\RequirePackage{lineno}
\documentclass[aps,prl,onecolumn,reprint,groupedaddress]{revtex4}
\usepackage{amsfonts,eucal,amsmath,amsthm,amssymb} 
\usepackage{graphicx}
\usepackage{color}
\usepackage{amsbsy}
\usepackage{blindtext}
\usepackage{lineno}
\usepackage{comment}

\def\a{{\bf a}}
\def\b{{\bf b}}

\def\dd{\mbox{d}}

\def\ve{\varepsilon}

\def\x{{\bf x}}

\newcommand{\veb}{\boldsymbol{\varepsilon}}

\begin{document}

\title{A kinetic theory for age-structured stochastic birth-death processes}

\author{Chris D. Greenman$^{1,2}$ and Tom Chou$^{3}$}
\affiliation{$^{1}$School of Computing Sciences, University of East
  Anglia, Norwich, UK, NR4 7TJ} 

\affiliation{$^{2}$The Genome Analysis Centre, Norwich Research Park,
  Norwich, UK, NR4 7UH} 

\affiliation{$^{3}$Depts. of Biomathematics and Mathematics, UCLA, Los
  Angeles, CA 90095-1766}





\begin{abstract}
Classical age-structured mass-action models such as the McKendrick-von
Foerster equation have been extensively studied but they are
structurally unable to describe stochastic fluctuations or
population-size-dependent birth and death rates. Stochastic theories
that treat semi-Markov age-dependent processes using \textit{e.g.},
the Bellman-Harris equation, do not resolve a population's
age-structure and are unable to quantify population-size
dependencies. Conversely, current theories that include size-dependent
population dynamics (\textit{e.g.}, mathematical models that include
carrying capacity such as the Logistic equation) cannot be easily
extended to take into account age-dependent birth and death rates. In
this paper, we present a systematic derivation of a new fully
stochastic kinetic theory for interacting age-structured
populations. By defining multiparticle probability density functions,
we derive a hierarchy of kinetic equations for the stochastic
evolution of an ageing population undergoing birth and death.  We show
that the fully stochastic age-dependent birth-death process precludes
factorization of the corresponding probability densities, which then
must be solved by using a BBGKY-like hierarchy.  However, explicit
solutions are derived in two simple limits and compared with their
corresponding mean-field results.  Our results generalize both
deterministic models and existing master equation approaches by
providing an intuitive and efficient way to simultaneously model age-
and population-dependent stochastic dynamics applicable to the study
of demography, stem cell dynamics, and disease evolution.

\end{abstract}

%

\maketitle


\section{Introduction}
Age is an important controlling feature in populations of
living organisms.  Processes such as birth, death, and mutation are
typically highly dependent upon an organism's chronological
age. Age-dependent population dynamics, where birth and death
probabilities depend on an organism's age, arise across diverse
research areas such as demography \cite{KEYFITZBOOK}, biofilm
formation \cite{AYATI2007}, and stem cell proliferation and
differentiation \cite{SUN2013,ROSHAN2014}. In this latter application,
not only does a the cell cycle give rise to age-dependent processes
\cite{QU2003,CELLCYCLE2014}, but the often small number of cells
requires a stochastic interpretation of the population.  Despite the
importance of age structure (such as that arising in the study of cell
cycles \cite{QU2003,CELLCYCLE2014,AGING0}), there exists no
theoretical method to fully quantify the stochastic dynamics of aging
and population-dependent processes. 

Past work on age-structured populations has focussed on {\it
  deterministic} models through the analysis of the so-called
McKendrick-von Foerster equation, first studied by McKendrick
\cite{MCKENDRICK,KEYFITZ} and subsequently von Foerster
\cite{VONFOERSTER}, Gurtin and MacCamy \cite{Gurtin1,Gurtin2}, and
others \cite{IANNELLI1995,WEBB2008}. In these classic treatments,
$\rho(a,t)\dd a$ is used to define, at time $t$, the density of
noninteracting agents with age between $a$ and $a+\dd a$.  The total
number of particles in the system at time $t$ is thus $n(t) =
\int_{0}^{\infty} \rho(a,t)\dd a$. If $\mu(a; n(t))$ is the death rate
for individuals of age $a$, the McKendrick-von Foerster equations are
\cite{Gurtin1,Gurtin2}

\begin{equation}
{\partial \rho(a,t)\over \partial t} + {\partial \rho(a,t)\over \partial a} = 
-\mu(a;n(t))\rho(a,t),
\label{MCKENDRICK0}
\end{equation}
with $\rho(a,t=0) = g(a)$ and

\begin{equation}
\rho(a=0,t) = \int_{0}^{\infty}
\beta(a;n(t))\rho(a,t)\dd a
\label{MCKENDRICKBC}
\end{equation}
for initial and boundary conditions, respectively. The boundary condition 
(Eq.~\ref{MCKENDRICKBC}) reflects the fact that birth gives rise to 
age-zero individuals. Note that the birth
and death rates $\beta$ and $\mu$ are usually simply assumed to be functions of the total
population $n(t)$.

The population dependence of $\beta(a;n(t))$ and $\mu(a;n(t))$ in
Eqs.~\ref{MCKENDRICK0} and \ref{MCKENDRICKBC} are assumed without
explicit derivation and it is not clear whether 
such simple expressions are self-consistent.  Moreover, the McKendrick-von Foerster equation
is expected to be accurate exact only when the dynamics of each
individual are not correlated with those of any other.  Therefore, a
formal derivation will allow a deeper understanding of how population
dependence and correlations arise in a fully stochastic age-structured
framework.

Two approaches that have been used for describing stochastic
populations include Master equations \cite{VANKAMPEN2011,FPTREVIEW}
and evolution equations for age-dependent branching process such as
the Bellman-Harris process
\cite{BELLMANHARRIS,REID1953,JAGERS1968,SHONKWILER1980,CHOUJTB}. Master-equation
approaches can be used to describe population-dependent birth or death
rates \cite{KENDALL1948,Gurtin1,Gurtin2,LJSALLEN} but implicitly
assume exponentially distributed waiting times between events
\cite{FPTREVIEW}. On the other hand, age-dependent models such as the
Bellman-Harris branching process \cite{BELLMANHARRIS} allow for
arbitrary distributions of times between birth/death events but they cannot
resolve age-structure of the entirte population nor 
describe population-dependent dynamics that arise from \textit{e.g.,}
regulation or environmental carrying capacities.

A number of approaches attempt to incorporate ideas of
stochasticity and noise into age-dependent population models,
\cite{SUN2013,REID1953,THESIS1998,DIFFUSION2009,GETZ1984,COHEN,LESLIE1945,LESLIE1948}.
For example, stochasticity can be implemented by assuming a random
rate of advancing to the next age window (by {\it e.g.}, stochastic
harvesting \cite{GETZ1984,COHEN} or a fluctuating environment
\cite{LANDE1988,LANDE2005}). However, such models do not account for
the intrinsic stochasticity of the underlying birth-death process that
acts differently on individuals at each different age.  One
alternative approach might be to extend the mean-field, age-structured
McKendrick-von Foerster theory into the stochastic domain by
considering the evolution of $P(n(a);t)$, the probability density that
there are $n$ individuals within age window $[a,a+\dd a]$ at time $t$
\cite{SUN2013,POLLARD1966}. This approach is meaningful only if a
large number of individuals exist in each age window, in which case a
large system size van Kampen expansion within each age window can be
applied \cite{VANKAMPEN2011}. However, such an assumption is
inconsistent with the desired small-number stochastic description of
the system.

A mathematical theory that addresses the age-dependent problem of
constrained stochastic populations would provide an important tool for
quantitatively investigating problems in demography,
bacterial growth, population biology, and stem cell differentiation and
proliferation. In this paper, we develop a new kinetic equation that
intuitively integrates population stochasticity, age-dependent effects
(such as cell cycle), and population regulation into a unified
theory. Our equations form a hierarchy analogous to that derived for
the BBGKY (Bogoliubov-Born-Green-Kirkwood-Yvon) hierarchy in kinetic
theory \cite{MCQUARRIE,ZANETTE}, allowing for a fully stochastic
treatment of age-dependent process undergoing population-dependent
birth and death.
%
%
%
%
%
%
%
\section{Kinetic equations for aging populations}
To develop a fully stochastic theory for age-structured populations
that can naturally describe both age- and population size-dependent
birth and death rates, we invoke multiple-particle distribution
functions such as those used in kinetic theories of gases
\cite{ZANETTE}.  Our analysis builds on the Boltzmann kinetic theory of
D. Zanette and yields a BBGKY-like hierarchy of equations. Here, the
positions of ballistic particles will represent the ages of 
individuals.

Changes in the total population require that we consider a family of
multiparticle distribution functions, each with different
dimensionality corresponding to the number of individuals. In this
picture, birth and death are represented by transitions between the
different distribution functions residing on different fixed
particle-number ``manifolds."  Processes that generate newborns
(particles of age zero) manifest themselves mathematically through
boundary conditions on higher dimensional distribution functions.

To begin, we define

\begin{equation}
f_{n}(x_{1}, x_{2},x_{3}, \ldots, x_{n}; t)\dd x_{1}\dd x_{2}\ldots\dd x_{n}
\end{equation}
as the probability that at time $t$, one observes $n$ distinguishable
(by virtue of their order of birth) individuals, such that the youngest
one has age within $(x_{1}, x_{1}+\dd x_{1})$, the second
youngest has age within $(x_{2}, x_{2}+\dd x_{2})$, and so on.
If the individuals are identical (except for their ages) and one does
not distinguish which are in each age window, one can define
$\rho_{n}(x_{1}, x_{2}, x_{3}, \ldots,x_{n};t)\dd x_{1}\dd
x_{2}\ldots\dd x_{n}$ as the probability that after randomly selecting
individuals, the first one chosen has age in $(x_1,x_{1}+\dd x_{1})$,
the second has age in $(x_{2}, x_{2}+\dd x_{2})$, and so on. For
example, if there are three individuals with ordered ages $x_1 < x_2 <
x_3$, the probability of making any specific random selection, such as
choosing the individual with age $x_{2}$ first, the one with age
$x_{1}$ second, and the one with age $x_{3}$ third, is
$\frac{1}{3!}$. More generally, when the ages $\x_{1,n}\equiv \x_{n} =
(x_1,x_2,\hdots, x_{n})$ are unordered, the associated probability
density is

\begin{equation}
\rho_n(\x_{n};t)=\frac{1}{n!}f_n({\cal T}(\{x_{i}\});t),
\label{ORDERFN}
\end{equation}
in which ${\cal T}$ is the time-ordering permutation operator such that,
for example, ${\cal T}(x_2,x_1,x_3)=(x_1,x_2,x_3)$.  Note that in this
formulation, $\rho_n(\x_{n};t)$ is invariant under interchange of the
elements of $\x_{n}$.

To derive kinetic equations for $\rho_n(\x_{n};t)$, we first define an
ordered cumulative probability distribution

\begin{equation}
Q_{n}(\a_{n};t) = 
\int_{0}^{a_{1}}\dd x_{1}\int_{x_{1}}^{a_{2}}\dd x_{2}\cdots
\int_{x_{n-1}}^{a_{n}}\!\!\!\dd x_{n} f_{n}(\x_{n};t),
\end{equation}
where $\a_{n} = a_{1,n} = (a_{1},\ldots,a_{n})$.  $Q_{n}(\a_{n};t)$
describes the probability that there are $n$ existing individuals at
time $t$ and that the youngest individual has age $x_{1}$ less than or
equal to $a_{1}$, the second youngest individual has age $x_{1}\leq
x_{2} < a_{2}$, and so on. The oldest individual has age $x_{n-1} \leq
x_{n} \leq a_{n}$.

\begin{figure}[t]
\begin{center}
\includegraphics[width=4.2in]{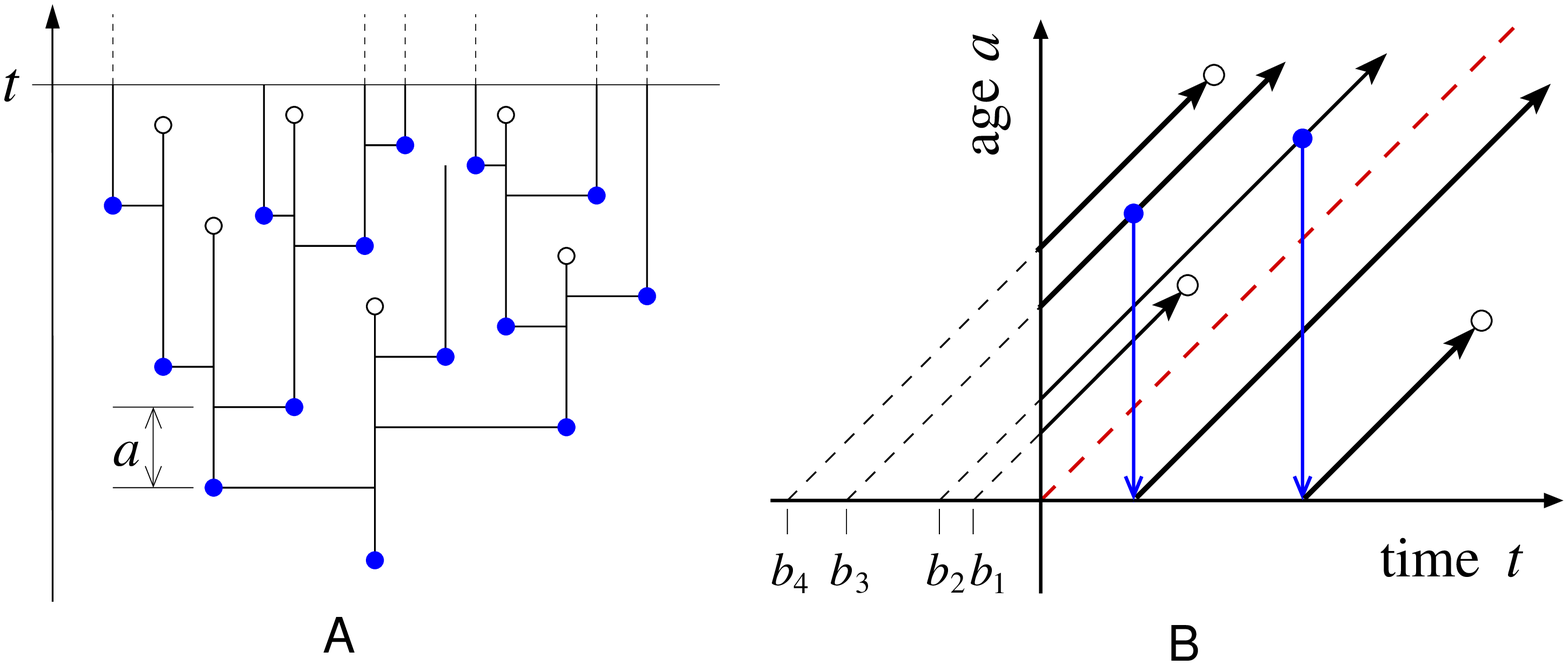}
\caption{\label{MODELS}(A) A simple age-dependent birth-death
  process. Each parent gives birth with an age-dependent rate
  $\beta_{n}(a)$, which may also depend on the total population size
  $n$.  Individuals can also die (open circles) at an age- and
  population-dependent rate $\mu_{n}(a)$. (B) Age trajectories in the
  upper ($a>t$) octant are connected to those in the lower one ($a<t$)
  through the birth processes.  Individuals that exist at time $t=0$
  can be traced back and defined by their time of birth $b_{i}$. Here,
  the labeling ordered according to increasing age. The pictured
  trajectories define characteristics $a_{i}(t)$ that can be used to
  solve Eq.~\ref{F0}.}
\end{center}
\end{figure}

We now compute the change in $Q_{n}(\a_{n}; t)$ over a small time
increment $\ve$: $Q_{n}(\a_{n} + \ve; t+\ve) = Q_{n}(\a_{n}; t) +
\int_{t}^{t+\ve}J(\a_{n};t')\dd t'$, where $J(\a_{n};t') =
J^{+}(\a_{n};t') - J^{-}(\a_{n};t')$ is the net probability flux at time
$t'$. The probability flux which increases the cumulative probability
is denoted $J^{+}$ while that which decrease the cumulative
probability is labelled $J^{-}$.  Each of the $J^{\pm}$ include
contributions from different processes that remove or add
individuals.  A schematic of our birth-death process,
starting from a single parent, is depicted in Fig.~\ref{MODELS}A.

In the $\ve \to 0$ limit, we find the conservation
equation

\begin{equation}
{\partial Q_{n}(\a_{n};t)\over \partial t}  + 
\sum_{i=1}^{n} {\partial Q_{n}(\a_{n};t)\over \partial a_{i}} = 
J^{+}(\a_{n};t)- J^{-}(\a_{n};t).
\label{Q0}
\end{equation}
Eq.~\ref{Q0} is a ``weak form'' integral equation for the probability density
which allows us to systematically derive an evolution equation and the
associated boundary conditions for $f_{n}(\x_{n};t)$.
The probability fluxes can be decomposed into components representing
age-dependent birth and death


\begin{equation}
J^{\pm}(\a_{n};t) = J^{\pm}_{\beta}(\a_{n};t) + J^{\pm}_{\mu}(\a_{n};t),
\label{JPM}
\end{equation}
where the birth and death that reduce probability 
can be expressed as

\begin{align}
\displaystyle J^{-}_{\beta}(\a_{n};t) =  \displaystyle
\int_{0}^{a_{1}}\!\!\dd x_{1}
\int_{x_{1}}^{a_{2}}\!\!\dd x_{2}
\cdots\!\int_{x_{n-1}}^{a_{n}}\!\!\!\!\!\dd x_{n}f_{n}(\x_{n};t)\sum_{i=1}^{n}\beta_{n}(x_{i}), \label{JMINUS_B}\\
\displaystyle J^{-}_{\mu}(\a_{n};t)  =   \displaystyle
\int_{0}^{a_{1}}\!\!\dd x_{1}
\int_{x_{1}}^{a_{2}}\!\!\dd x_{2}
\cdots\!\int_{x_{n-1}}^{a_{n}}\!\!\!\!\!\dd x_{n}f_{n}(\x_{n};t)\sum_{i=1}^{n}
\mu_{n}(x_{i}).\label{JMINUS_MU}
\end{align}
Similarly, the probability fluxes that increase probability 
are

\begin{align}
\displaystyle J^{+}_{\beta}(\a_{n};t) = &
\int_{0}^{a_{2}}\!\!\dd x_{1}\cdots\!
\int_{x_{j-1}}^{a_{j+1}}\!\!\!\dd x_{j}\cdots\!\int_{x_{n-2}}^{a_{n}}\!\!\!\!\dd x_{n-1}
f_{n-1}(\x_{n-1};t)\sum_{i=1}^{n-1}\beta_{n-1}(x_{i}), \label{JPLUS_B}
\end{align}
\begin{align}
\displaystyle J^{+}_{\mu}(\a_{n};t) \displaystyle = & \sum_{i=0}^{n}
\int_{0}^{a_{1}}\!\!\dd x_{1}\cdots\!\int_{x_{i-1}}^{a_{i}}\!\!\dd x_{i}
\int_{x_{i}}^{a_{i+1}}\!\!\!\dd y \int_{y}^{a_{i+1}}\!\!\!\!\dd x_{i+1} \cdots\!
\int_{x_{n-1}}^{a_{n}}\!\!\!\dd x_{n}\,\mu_{n+1}(y) 
f_{n+1}(\x_{i},y,\x_{i+1,n};t),\label{JPLUS_MU}
\end{align}
in which $\x_{i,j}\equiv (x_{i},x_{i+1},\ldots,x_{j}), x_{0} \equiv
0$, $a_{n+1}\equiv \infty$, and the age- and population-dependent
birth and death rates for individual $i$ are denoted
$\beta_{n}(x_{i})$ and $\mu_{n}(x_{i})$, respectively.  The
probability flux into $Q_{n}(\a_{n};t)$ arising from birth of the
$n-1$ individuals of age $\a_{2,n}\equiv (a_{2}, a_{3},\ldots,a_{n})$
generates an individual of age zero. Hence, a key feature of
$J_{\beta}^{+}(\a_{n};t)$ is that it does not 
depend on $a_{1}$. 

We can now describe the fully stochastic aging process in terms of the
ordered distribution function $f_{n}(\x_{n};t)$ by using Eqs.~\ref{JPM}-\ref{JPLUS_MU}
in Eq.~\ref{Q0} and applying the operator ${\partial
\over \partial a_{n}}\cdots{\partial \over \partial a_{2}}{\partial
\over \partial a_{1}}$ to find

\begin{equation}
\begin{array}{l}
\displaystyle {\partial f_{n}(\a_{n};t)\over \partial t} + \sum_{j=1}^{n}
{\partial f_{n}(\a_{n};t)\over \partial a_{j}} = 
-f_{n}(\a_{n};t)\sum_{i=1}^{n} \gamma_{n}(a_{i})+ \sum_{i=0}^{n}\int_{a_{i}}^{a_{i+1}}\!\!\!\!
\mu_{n+1}(y)f_{n+1}(\a_{i},y,\a_{i+1,n};t)\dd y,
\label{F0}
\end{array}
\end{equation}
where $a_{0}\equiv 0$, $a_{n+1}\equiv \infty$, and the total age-dependent transition rate is
\begin{equation}
\gamma_{n}(a_{i}) = \beta_{n}(a_{i})+ \mu_{n}(a_{i}).
\end{equation}
Note that the $a_{1}-$independent source term $J^{+}_{\beta}$ that had
contributed to the ordered cumulative (Eq.~\ref{Q0}) does not
contribute to the bulk equation for $f_{n}(\a_{n};t)$. Rather, it
arises in the boundary condition for $f_{n}$, which can be found by
setting $a_{1} = 0$ in Eq.~\ref{Q0}. Since $Q(0,a_{2}, \ldots,
a_{n};t) = 0$ and $J_{\beta}^{+}(\a_{n};t)$ are
independent of $a_{1}$, the remaining terms are

\begin{equation}
\int_{0}^{a_{2}}\!\!\dd x_{2}\cdots\int_{x_{n-1}}^{a_{n}}\!\!\!\!
\dd x_{n} f_{n}(x_{1}=0,\x_{2,n};t) = 
J^{+}_{\beta}(\a_{n};t).
\label{BCQ}
\end{equation}
Further taking the derivatives ${\partial \over \partial
  a_{n}}\cdots{\partial \over \partial a_{2}}$ of Eq.~\ref{BCQ}, we
find the boundary condition

\begin{equation}
f_{n}(a_{1}=0,\a_{2,n};t) = 
f_{n-1}(\a_{2,n};t)\sum_{i=2}^{n} \beta_{n-1}(a_{i}).
\label{BCF}
\end{equation}
%

We now consider {\it indistinguishable} individuals as described by
the density defined in Eq.~\ref{ORDERFN}. Equation \ref{F0} can then
be expressed in terms of $\rho_{n}(\a_{n};t)$: the probability density
that if we {\it randomly} label individuals, the first one has age
between $a_{1}$ and $a_{1}+\dd a_{1}$, the second has age between
$a_{2}$ and $a_{2}+\dd a_{2}$, and so on.  The kinetic equation for
$\rho_{n}$ can then be expressed in the form

\begin{align}
\displaystyle {\partial \rho_{n}(\a_{n};t)\over \partial t} + 
& \sum_{j=1}^{n}{\partial \rho_{n}(\a_{n};t)\over \partial a_{j}} = 
-\rho_{n}(\a_{n};t)\sum_{i=1}^{n} \gamma_{n}(a_{i}) 
 + (n+1)\!\int_{0}^{\infty}\!\!\mu_{n+1}(y)
\rho_{n+1}(\a_{n},y;t)\dd y, \label{RHO0} 
\end{align}
and the boundary condition becomes

\begin{equation}
\begin{array}{l}
n \rho_{n}(a_{1},\ldots,a_{\ell}=0,\ldots, a_{n};t) = \rho_{n-1}(a_{1},\ldots,
\hat{a}_{\ell},\ldots,a_{n};t)\sum_{i(\ne \ell)=1}^{n}\beta_{n-1}(a_{i}),
\label{BCRHO}
\end{array}
\end{equation}
where the sum precludes the $i=\ell$ term and $\hat{a}_{\ell}$
indicates that the variable $a_{\ell}$ is omitted from the sequence of
arguments \cite{ZANETTE}. Equation \ref{RHO0} and the boundary
conditions of Eq.~\ref{BCRHO}, along with an initial condition
$\rho_{n}(\a_{n};t=0)$, fully define the stochastic age-structured
birth-death process and is one of our main results.  Eq.~\ref{RHO0} is
analogous to a generalized Boltzmann equation for $n$ particles
\cite{ZANETTE,PETERS}.  The evolution operator corresponds to that of
free ballistic motion in one dimension corresponding to age.  However,
instead of particle collisions typically studied in traditional
applications of the Boltzmann equation, our problem couples density
functions for $n$ particles to those of $n+1$ and $n-1$ (through the
boundary condition).
\section{Solutions and equation hierarchies}
Equation \ref{RHO0} defines a set of coupled linear
integro-differential equations. We would like to find solutions for
$\rho_n(\a_n;t)$ expressed in terms of an initial condition
$g_{n}(\a_{n}-t;t=0)$. However, we will see below that the presence of
births during the time interval $[0,t]$ prevents a simple solution to
Eq. \ref{RHO0} due to interference from the boundary condition in
Eq. \ref{BCRHO}. Instead, we will obtain a solution for
$\rho_{n}(\a_{n};t)$ at time $t$ in terms of the distribution
$\rho_{n}(\a_{n}-(t-t_0);t_0)$ at an earlier time $t_0$ selected such
that no births occur during the time interval $(t_0,t]$. That is, if
  $b_i=t-a_i$ represents the time of birth of the $i^{\rm th}$
  individual (see Fig.~\ref{MODELS}B), we have the condition $t_0 \ge
  b_{i} \,\forall\, i$. The dynamics described by Eq.~\ref{RHO0} are
  then unaffected by the boundary condition (Eq.~\ref{BCRHO}) and can
  be solved using the characteristics $a_i = t - b_i$ indexed by
  individual times of birth $b_{i}$.  Note that any individual
  initially present (at time $t=0$) has a projected negative time of
  birth. We can then solve $\rho_{n}(t-\b_{n};t)$ explicitly along
  each characteristic and then re-express them in terms of $\a_{n}$,
  to obtain
\begin{align}
\rho_{n}(\a_{n};t)&  =  U_{n}(\a_{n};t_0;t)\rho_{n}(\a_{n}-(t-t_0);t_0)
 +(n+1)\int_{t_0}^{t}U_{n}(\a_{n};t';t)
\left[\int_{0}^{\infty}\!\!\!\mu_{n+1}(y)
\rho_{n+1}\dd y\right]\dd t',\label{SOLN0}
\end{align}
where $\rho_{n+1} \equiv \rho_{n+1}(\a_{n}-(t-t'),y;t')$ above, and 
\begin{align}
U_{n}(\a_{m};t';t) & = 
\exp\left[-\sum_{i=1}^{m}\int_{t'}^{t}\!\gamma_{n}(a_{i}-(t-s))\dd s\right]
\equiv U_{n}^{-1}(\a_{m};t_0;t')U_{n}(\a_m;t_0;t)
\label{PROP}
\end{align}
is the propagator for any set of $m \leq n$ 
individuals from time $t'$ to $t$. 

In the case of a pure death process where no births occur ($\beta_{n}=0$),
allowing us to set $t_0=0$. A complete solution can
be found through successive iteration of Eq.~\ref{SOLN0}. We further simplify
matters by assuming an initial condition that factorizes into an
initial total number distribution $\rho(n)$ and common initial age
probability densities $g(a)$:
$\rho_n(\a_n-t;0)=\rho(n)\prod\limits_{i=1}^{n}g(a_i-t)$.  If we further
assume a death rate $\mu_n(a)=\mu(a)$ that is independent of
population size, Eq.~\ref{SOLN0} can be solved, after some algebra, to
yield

\begin{align}
\rho_n(\a_n;t) = & U(\a_n;0;t)\prod\limits_{i=1}^{n} g(a_i-t)
\displaystyle\sum\limits_{k=0}^\infty{n+k \choose k}\rho(n+k) \left[\int\limits_0^{t} g(y-s)\int\limits_s^{\infty} U(y;0;s)\mu(y) \dd y \dd s \right]^k.
\label{DEATHEQNSOL}
\end{align}

For a pure birth process where $\mu_{n}=0$, the second integral term
in Eq.~\ref{SOLN0} disappears.  In this case, we must use the boundary
condition (Eq.~\ref{BCRHO}) to successively bootstrap the solution by
applying the propagator $U$ between birth times.  Assume a starting
time $t=0$ with an initial condition consisting of $m$ individuals
with corresponding ages $a >t$.  The symmetry of 
$\rho_n(\a_n;t)$ and $U_n(\a_n;t';t)$ implies that, without
loss of generality, ages can be arranged in decreasing order: $a_1 > a_2 >
\hdots > a_{m} > t > a_{m+1} > \hdots > a_n$, where the youngest was
born most recently at time $t-a_n>0$.  If we select $t_0$ to be the
moment of birth at time $b_n=t-a_{n}$ of the most recently born
($n^{\rm th}$) individual, the density over all individuals is
propagated forward according to

\begin{equation}
\rho_{n}(\a_{n};t) = U_{n}(\a_{n};b_{n};t)\rho_{n}(\{\a_{n-1}-a_{n},0\};t-a_{n}),
\end{equation}
where $\rho_{n}(\{\a_{n-1}-a_{n},0\};t-a_{n})$ is the initial condition immediately after the 
birth of the $n^{\rm th}$ individual and can be related to $\rho_{n-1}$ through the 
boundary condition in Eq.~\ref{BCRHO}. The density function thus obeys
\begin{align}
\rho_{n}(\a_{n};t) = & \frac{1}{n}U_{n}(\a_{n};b_{n};t)
\rho_{n-1}(\a_{n-1}-a_{n};t-a_{n})\sum_{i=1}^{n-1}\beta_{n-1}(a_{i}-a_{n}).
\label{ITERATION}
\end{align}
Eq.~\ref{ITERATION} can then be iterated back to $t=0$ to find the
solution for randomly selected individuals. For the case in which
$\gamma_{n} = \gamma$ is independent of the population size, the
propagator can be separated into a product across individuals. If
$\beta_{n} = \beta$ is also independent of $n$, the solution takes the
simple form

\begin{align}
\rho_{n}(\a_{n};t) = & g_{m}(\a_{m}-t) U(\a_{m};0;t)
{m! \over n!} \!\!\!\prod_{k=m+1}^{n}U
(a_k;b_k;t)
\sum_{\ell=1}^{k-1}\beta(a_{\ell}-a_{k}),
\label{BIRTHSOL}
\end{align}
where $b_{k} = t-a_{k}$ and  $g_{m}$ is the initial distribution 
of ages for the $m$ individuals born before $t=0$.

The above solutions for $\rho_{n}(\a_{n};t)$ allow us to
explicitly compare differences between the fully stochastic theory and
the deterministic McKendrick-von Foerster model.  As an example,
consider the expected number of individuals at time $t$ that have age
between $0$ and $a$,

\begin{equation}
P(a,t) = \int_{0}^{a}\!\!\rho(y,t)\dd y,
\label{PAT}
\end{equation}
where $\rho(y,t)$ is found from Eqs.~\ref{MCKENDRICK0} and
\ref{MCKENDRICKBC}.  We wish to compare this quantity with the {\it
  probability} $P_m(a,t)$ that there are $m$ individuals at time $t$
with age between $0$ and $a$.  The probability $P_m(n,a,t)$ that there
are $n$ total individuals of which exactly $m$ have age between $0$
and $a$ can be constructed from our fully stochastic theory via

\begin{equation}
P_m(n,a,t) = {n \choose m}\prod_{j=1}^{m}\int_{0}^{a}\dd a_{j}\!\!\!
\prod_{\ell=m+1}^{n}\!\!\!\int_{a}^{\infty}\!\!\!\!\dd a_{\ell}\, \rho_{n}(\a_{n};t).
\label{PMNAT}
\end{equation}
The marginal probability $P_m(a,t)$ is then found by summing over $n\geq m$:

\begin{equation}
P_m(a,t) = \sum_{n=m}^{\infty}P_m(n,a,t).
\label{STOCHCUMUL}
\end{equation}
The comparison can be made more explicit by considering simple cases
such as an age-independent birth-only process with fixed birth rate
$\beta$. If the process starts with precisely $N$ individuals,
standard methods \cite{IANNELLI1995,WEBB2008} yields a simple solution
of the McKendrick-von Foerster equation which when used in
Eq.~\ref{PAT} gives $P(a<t;t)=Ne^{\beta t}\left(1-e^{-\beta
  a}\right)$. Substituting the pure birth solution of
Eq.~\ref{BIRTHSOL} into Eqs.~\ref{PMNAT} and \ref{STOCHCUMUL} yields
\begin{equation}
P_m(a,t)={m+N-1 \choose m}\frac{e^{-N\beta t}\left(1-e^{-\beta a}\right)^m}
{\left(1-e^{-\beta a}+e^{-\beta t}\right)^{m+N}}.
\end{equation}
\begin{figure}[t]
\begin{center}
\includegraphics[width=2.1in]{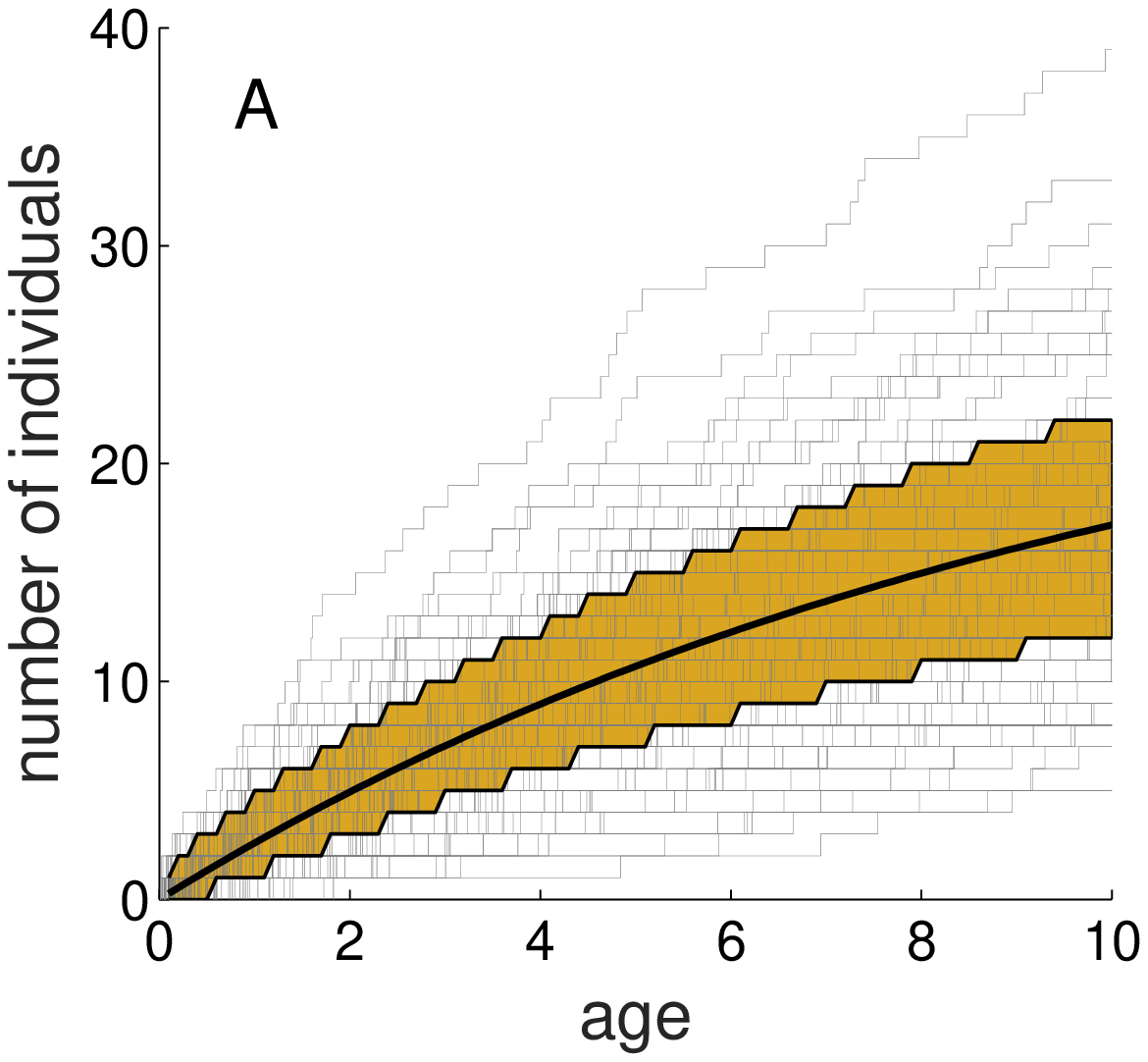}\hspace{1.3cm}\includegraphics[width=2.1in]{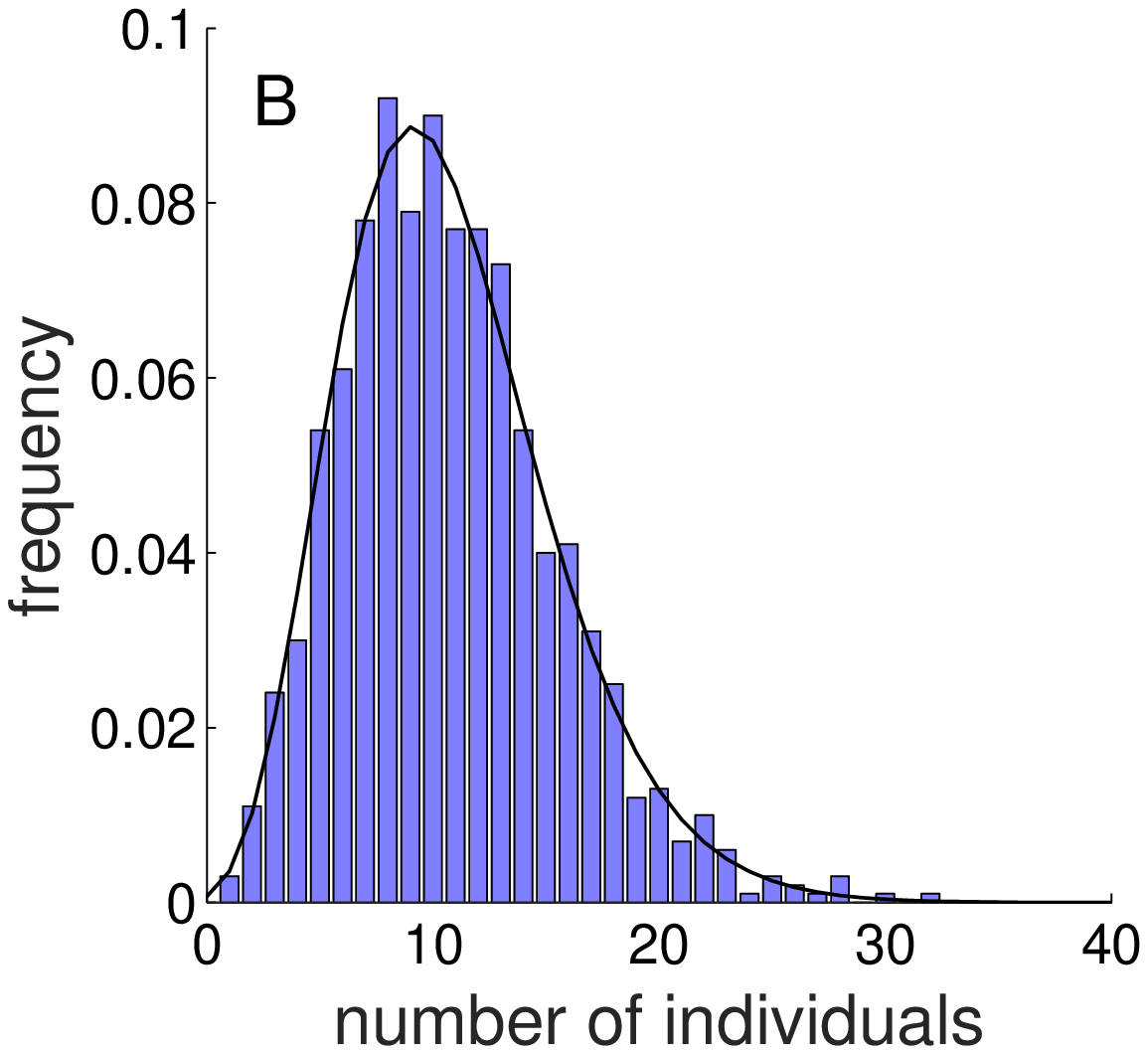}
\caption{\label{Birth_Example}Comparison of $P(a,t)$ (Eq.~\ref{PAT})
  derived from the McKendrick-von Foerster equation with $P_{m}(a,t)$
  of a fully stochastic pure birth process with constant
  $\beta=0.1$. We start with $N=10$ individuals and analyze our
  quantities at time $t=10$ for ages $a<t$.  (A) Each of the 100 grey
  lines count the number of individuals younger than age $a$ in one
  simulation. The solid black curve indicates the deterministic
  (McKendrick-von Foerster) solution $P(a,t)=\int_0^a \rho(y,t)\dd y$,
  which can also be obtained through
  $P(a,t)=\sum_{m=1}^{\infty}mP_{m}(a,t)$. The shaded region
  represents the inter-quartile range of $P_{m}(a,t)$. (B)
  Distribution constructed from 1000 simulations (bars) and
  theoretical distribution $P_m(a=5,t=10)$ (black curve).}
\end{center}
\end{figure}

In Fig.~\ref{Birth_Example}A we compare the expected value $P(a,t)$
derived from solutions to the McKendrick-von Foerster equation with
stochastic simulations that sample the stochastic result $P_m(a,t)$.
The fully stochastic nature of the process is clearly shown by the
spread of the population about the expected
value. Fig.~\ref{Birth_Example}B plots the corresponding number
distribution $P_{m}(5,10)$.


Finally, to connect our general kinetic theory with
statistically-reduced (and deterministic) descriptions, we consider
reduced $k-$dimensional distribution functions defined by integrating
$\rho_n(\a_n;t)$ over $n-k$ age variables:

\begin{equation}
\rho_{n}^{(k)}(\a_{k};t) \equiv \int_{0}^{\infty}\!\dd a_{k+1}
\ldots\int_{0}^{\infty}\!\dd a_{n}\, \rho_{n}(\a_{n};t).
\end{equation}
The symmetry properties of $\rho_{n}(\a_{n};t)$ indicate that it is
immaterial which of the $n-k$ age variables are integrated out.  If we
integrate Eq.~\ref{RHO0} over all ages ($k=0$), and assume
$\rho_{n}^{(1)}(a = \infty;t) = 0$, we find

\begin{align}
{\partial \rho_{n}^{(0)}(t) \over \partial t} = &  n\rho_{n}^{(1)}(a=0;t) 
-n \int_{0}^{\infty}\!\!\!\gamma_{n}(y)\rho_{n}^{(1)}(y;t)\dd y +(n+1)\int_{0}^{\infty}\!\!\!\mu_{n+1}(y)\rho_{n+1}^{(1)}(y;t)\dd y.
\label{ZERO0}
\end{align}
Furthermore, integrating Eq.~\ref{BCRHO} over $a_{i\neq \ell}$ 
yields  $n\rho_{n}^{(1)}(a=0;t) = (n-1)
\int_{0}^{\infty}\beta_{n-1}(y)\rho_{n-1}^{(1)}(y;t)\dd y$.
Thus, Eq.~\ref{ZERO0} can be written in the form

\begin{align}
\displaystyle {\partial \rho_{n}^{(0)}(t) \over \partial t} = & (n-1)\int_{0}^{\infty}
\beta_{n-1}(y)\rho_{n-1}^{(1)}(y;t)\dd y - n \int_{0}^{\infty}(\beta_{n}(y)+\mu_{n}(y))\rho_{n}^{(1)}(y;t)\dd y  +(n+1)\int_{0}^{\infty}\!\mu_{n+1}(y)\rho_{n+1}^{(1)}(y;t)\dd y.
\label{ZERO1}
\end{align}

Eq.~\ref{ZERO1} describes the evolution of the probability
$\rho_{n}^{(0)}(t)$ that the system contains $n$ individuals at time
$t$ and it contains the single-particle marginal density
$\rho_{n}^{(1)}(y;t)$. Upon deriving equations for
$\rho_{n}^{(1)}(y;t)$, one would find that they depend on
$\rho_{n}^{(2)}(y_{1}, y_{2};t)$, and so on.  Therefore, the marginal
probability densities form a hierarchy of equations, as is typically
seen in classic settings such as the kinetic theory of gases
\cite{MCQUARRIE} and the statistical theory of turbulence
\cite{FRISCH}. Note that if the birth and death rates $\beta_{n}$ and
$\mu_{n}$ are age-independent, they are constants with respect to the
integral and Eq.~\ref{ZERO1} reduces to the familiar constant birth and death rate
master equation for the simple birth-death process:

\begin{align}
{\partial \rho_{n}^{(0)}(t) \over \partial t} = & (n-1)\beta_{n-1}\rho_{n-1}^{(0)}(t)
-n(\beta_{n}+\mu_{n})\rho_{n}^{(0)}(t) + (n+1)\mu_{n+1}\rho_{n+1}^{(0)}(t),
\label{ZERO2}
\end{align}
where $\rho_{n}^{(0)}(t)$ is the probability the system contains $n$ 
individuals at time $t$, regardless of their ages.

In general, integration of Eq.~\ref{RHO0} over $n-k\geq 0$ age variables
leaves $k$ remaining independent age variables. The resulting kinetic
equation for $\rho_n^{(k)}(\a_{k}; t)$ involves both
$\rho_{n+1}^{(k+1)}(\a_{k},y;t)$ and boundary terms
$\rho_{n}^{(k+1)}(\a_{k},a_{k+1}=0;t)$. These boundary terms can be
eliminated by using the result obtained from integration of the
boundary condition (Eq.~\ref{BCRHO}) over $n-k-1$ age variables. By
exploiting the symmetry properties of the marginals $\rho_n^{(k)}$, we
find

\begin{align}
\displaystyle {\partial \rho_{n}^{(k)}(t) \over \partial t}   + 
\sum_{i=1}^{k}{\partial \rho_{n}^{(k)}(\a_{k};t)\over \partial a_{i}}  = &
+ \left({n-k \over n}\right)\rho_{n-1}^{(k)}(\a_{k};t)
\sum_{i=1}^{k}\beta_{n-1}(a_{i}) +{(n-k)(n-k-1)\over n}\int_{0}^{\infty}\beta_{n-1}(y)
\rho_{n-1}^{(k+1)}(\a_{k},y;t)\dd y \nonumber \\
\: & \displaystyle -\rho_{n}^{(k)}(\a_{k};t)\sum_{i=1}^{k}\gamma_{n}(a_{i})
-(n-k)\int_{0}^{\infty}\gamma_{n}(y)
\rho_{n}^{(k+1)}(\a_{k},y;t)\dd y \label{RHOK1} \\ 
\: & \displaystyle + (n+1)\int_{0}^{\infty}\mu_{n+1}(y)
\rho_{n+1}^{(k+1)}(\a_{k},y;t)\dd y.\nonumber
\label{RHOK1}
\end{align}
Each function $\rho_{n}^{(k)}$ in the hierarchy not only depends on
the functions in the $n \pm 1$ subspace, but is connected to
functions with $k+1$ and $k-1$ variables. The latter
coupling arises through the boundary condition for $\rho_{n}^{(k)}$ which
involves densities $\rho_{n}^{(k-1)}$. As with similar equations in
physics, the hierarchy of equations cannot be generally solved, and
either factorization approximations or truncation (such as moment
closure) must be used. 

We now show that the $k=1$ equation explicitly leads to the
classic McKendrick-von Foerster equation and its associated boundary
condition. For $k=1$, $\rho_{n}^{(1)}(a;t)\dd a$ is the probability
that there are $n$ individuals and that if one is randomly chosen, it
will have age between $a$ and $a+\dd a$. Therefore, the probability
that we have $n$ individuals of which any one has age between $a$ and
$a+\dd a$ is $n\rho_{n}^{(1)}(a; t)\dd a$. Summing over all possible
population sizes $n\geq 1$ gives us the probability $\rho(a,t)\dd a$ that the system
contains an individual with age between $a$ and $a+\dd a$:

\begin{equation}
\rho(a,t) \equiv \sum_{n=0}^{\infty}n\rho_{n}^{(1)}(a;t).
\end{equation}
Multiplying Eq.~\ref{RHOK1} (with $k=1$) by $n$ and summing 
over all positive integers $n$, we find after carefully cancelling like terms

%

\begin{equation}
\displaystyle {\partial \rho(a,t) \over \partial t} + 
{\partial \rho(a,t)\over \partial a} = -\sum_{n=1}^{\infty}n\mu_{n}(a)\rho_{n}^{(1)}(a;t).
\label{MVF0}
\end{equation}
Equation~\ref{MVF0} generalizes the McKendrick-von Foerster model 
to allow for population-dependent death rates,
but  does not reduce to the 
simple form shown in Eq.~\ref{MCKENDRICK0}. 
Population-dependent effects in equation for $\rho(a,t)$ requires 
requires knowing the ``single-particle'' density function
$\rho_{n}^{(1)}(a;t)$ and subsequently all higher 
order distribution functions.

A boundary condition is naturally recovered by integrating over all
ages but $a_{\ell}$ in Eq.~\ref{BCRHO} and summing over all $n$:
\begin{align}
\sum_{n=1}^{\infty}n\rho_{n}^{(1)}(a=0;t) & \equiv \rho(a=0,t) = \sum_{n=2}^{\infty}(n-1)
\!\int_{0}^{\infty}\!\!\!\beta_{n-1}(y)\rho_{n-1}^{(1)}(y;t)\dd y.
\label{BCMVF0}
\end{align}
These equations show that the McKendrick-von Foerster equation is
recovered only if both $\mu_{n}(a) = \mu(a)$ and $\beta_{n}(a) =
\beta(a)$ are independent of population size. In this case, $\mu(a)$ can be
pulled out of the sum in Eq.~\ref{MVF0} and
$\sum_{n=1}^{\infty}n\mu_{n}(a)\rho_{n}^{(1)}(a;t)=
\mu(a)\rho(a,t)$. Similarly,
$\int_{0}^{\infty}\beta(y)\left[\sum_{n=2}^{\infty}(n-1)
  \rho_{n-1}^{(1)}(y;t)\right]\dd y =
\int_{0}^{\infty}\beta(y)\rho(y,t)\dd y$, which is the simple boundary
condition associated with the classic McKendrick-von Foerster
model. This derivation clearly shows that population-dependent birth
and death rates cannot be readily incorporated into an age-dependent
model, even one that is deterministic, without considering the
hierarchy of population densities.
\section{Discussion and conclusions}
We have developed a complete kinetic theory for age-structured
birth-death processes.  To stochastically describe the age structure
of a population requires a higher dimensional probability density. The
evolution of this high-dimensional probability density mirrors that
found in the Boltzmann equation for one-dimensional, ballistic,
noninteracting gas dynamics. However, one crucial difference is that
the number of individuals can increase or decrease according
to the age-dependent birth and death rates.  Thus, the dynamics are
determined by a phase-space-conserving Liouville operator so long as
the number of individuals does not change \cite{MCQUARRIE}. Once an
individual is born or dies, the system jumps to another manifold in a
higher or lower dimensional phase-space, immediately after which
conserved dynamics resume until the next birth or death event.  Such
variable number dynamics share similarities with the kinetic theory of
chemically reacting gases \cite{REACTING0}.

Our main mathematical results are Eqs.~\ref{RHO0} and
\ref{BCRHO}. These equations show that birth-death
dynamics couple densities associated with different numbers $n$ and
describes the process in terms of ballistically moving particles all
moving with unit velocity in the age ``direction.'' The individual
particles can die at rates that depend on their distance from their
origin (birth). Particles can also give birth at rates dependent on
their age. The injection of newborns at the origin (zero age) is
described by the boundary condition (Eq.~\ref{BCRHO}).

One important advantage of our approach is that it provides a natural
framework for incorporating both age- and population-dependent birth
and death rates into a stochastic description, which has thus far not
been possible with other approaches.  In general, our kinetic
equations need to be solved numerically; however, we found analytic
expressions for $\rho_{n}(\a_{n};t)$ when either birth or death
vanishes and the other is independent of population. Furthermore, we
define marginal density functions and develop a hierarchy of equations
analogous to the BBGKY hierarchy (Eq.~\ref{RHOK1}).  These equations
for the marginal densities allow one to construct any desired
statistical measure of the process and are also part of our main
results.  We explicitly showed how a zeroth order equation leads to
the equation for the marginal probability of observing $n$ individuals
in the standard {\it age-independent} birth-death processes
(Eq.~\ref{ZERO2}) \cite{LJSALLEN}.  The first-order equation is also
used to derive a hybrid equation for the mean density $\rho(a,t)$ that
involves the single-particle density function $\rho_{n}^{(1)}(a;t)$
(which ultimately depends on higher-dimensional densities
$\rho_{n}^{(k>1)}(\a_{k};t)$ through the hierarchy). Only when death
is independent of population does the theory reduce to the
deterministic McKendrick-von Foerster equation (Eq.~\ref{MVF0}) and
the associated boundary condition (Eq.~\ref{BCMVF0}). 

Extensions of our high-dimensional age-structured kinetic theory to
more complex birth-death mechanisms such as sexual reproduction and
renewal/branching processes can be straightforwardly investigated. The
simple birth-death process we analyzed allows for the birth of only a
single age-zero daughter from a parent at a time. We note that the
Bellman-Harris process described via generating functions
\cite{JAGERS1968,SHONKWILER1980} (which can describe age-dependent
death and branching, but cannot be used to model population-dependent
dynamics) assumes self-renewal at each branching event. That is, two
(or more) daughters of zero age are simultaneously produced from a
parent.  Such differences in the underlying birth process can lead to
qualitative differences in important statistical measures beyond
mean-field, such as first passage times \cite{CHOUJTB}.  The
branching/renewal process, as well as sexual reproduction, requires
nontrivial extensions of our kinetic theory and will be explored in a
future investigation.
\section{Acknowledgements}
This research was supported in part at KITP by the National Science
Foundation under Grant No. PHY11-25915. TC is also supported by the 
NIH through grant R56 HL126544 and the Army Research Office through grant 
W911NF-14-1-0472.

\bibliography{refs1}

\end{document}